# ASURA: Scalable and Uniform Data Distribution Algorithm for Storage Clusters

Ken-ichiro Ishikawa, System Platform Research Laboratories, NEC Corporation

**Abstract**—Large-scale storage cluster systems need to manage a vast amount of data locations. A naïve data locations management maintains pairs of data ID and nodes storing the data in tables. However, it is not practical when the number of pairs is too large. To solve this problem, management using data distribution algorithms, rather than management using tables, has been proposed in recent research. It can distribute data by determining the node for storing the data based on the datum ID. Such data distribution algorithms require the ability to handle the addition or removal of nodes, short calculation time and uniform data distribution in the capacity of each node. This paper proposes a data distribution algorithm called ASURA (Advanced Scalable and Uniform storage by Random number Algorithm) that satisfies these requirements. It achieves following four characteristics: 1) minimum data movement to maintain data distribution according to node capacity when nodes are added or removed, even if data are replicated, 2) roughly sub-micro-seconds calculation time, 3) much lower than 1% maximum variability between nodes in data distribution, and 4) data distribution according to the capacity of each node. The evaluation results show that ASURA is qualitatively and quantitatively competitive against major data distribution algorithms such as Consistent Hashing, Weighted Rendezvous Hashing and Random Slicing. The comparison results show benefits of each algorithm; they show that ASURA has advantage in large scale-out storage clusters.

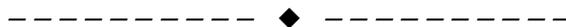

## 1 INTRODUCTION

THE growth of data that need to be managed by computers has been increasing the capacity of storage systems. The capacities in the recent storage system cannot be achieved with one or a few storage nodes. Thus, managing technologies to group many storage nodes as a single storage system are urgently required.

There are two types of managing technologies: storage cluster technology and distributed storage system (peer-to-peer (P2P) system). In the storage cluster technology, each node in a storage system knows all the nodes. On the other hand, in the distributed storage system, each node in a storage system knows only some of the nodes. This paper focuses on the storage cluster technology.

When data are accessed, the nodes storing them need to be determined based on their datum ID. Therefore, all pairs of data IDs and data-storing nodes (the node that stores the datum) must be managed.

There are three types of storage cluster technologies: table-based management, algorithm-based management, and a combination of both.

The table-based management memorizes pairs of data IDs and data-storing nodes in a management table. When a datum is to be accessed, the corresponding datum ID is searched for in the management table, and the ID offers access to the corresponding node. An example of the table-based management system is the Google File System [1] and Hadoop Distributed File System [46]. The table-based management is easy to implement, but it requires a large table if the data size is large. Sharing and synchronizing such a large table also consumes memory and network bandwidth. Furthermore, if the table is known only by management nodes, they are accessed by all nodes. Thus, the management nodes become a bottleneck.

In contrast to the table-based management, the algorithm-based management, which calculates a data-storing node from a datum ID, does not need to manage and synchronize such a large table. Examples of algorithm-based management systems are Dynamo [2] and Ceph [3]. The difficulty in the algorithm-based management lies in algorithm design. The algorithm must be able to determine the node corresponding to any given datum ID. It must also offer the ability to handle the addition or removal of nodes, low resource usage, and good load balancing according to capacity of each node. Since it does not need to memorize data-node combinations, it only requires a small table to memorize node information, and it only needs to share this small table among all the nodes that determine which nodes are data-storing nodes.

The size of the table in the table-based management approach is too large for practical use with large-scale storage clusters. Therefore, management using an algorithm, whether totally or partially, is the only choice for huge storage.

One of the main applications of the algorithm-based management is large scale-out storage clusters. The author think that following characteristics are desirable for algorithms for large scale-out storage clusters. The first characteristic is that only a small number of data movements are required to maintain appropriate data distribution among nodes, when nodes are added or removed. In particular, it is preferable if the number of data movement is minimum. The second characteristic is short calculation time. Since storage clusters communicate through networks, calculation times need to be shorter enough than the network communication time. The third characteristic is uniform data distribution, which offers both full use of all nodes and good load balancing. In addition, it is preferable that data are distributed according to the capacity of each node. The last characteristic is the ability to support data replication, because large scale-out storage clusters must store replicated data to achieve failure recovery.

---

• *Ken-ichiro Ishikawa is with NEC Corporation, Kawasaki, Japan E-mail: ishikawa@kikurage.net*

This paper focuses on a class of algorithm that satisfies minimum data movement when nodes are added or removed, uniform data distribution, and scalability. These algorithms use pseudorandom functions to achieve pseudo-equal data distribution. The first example of this class of algorithms is Consistent Hashing [4]. The second is Weighted Rendezvous Hashing [5], and the third is Random Slicing [6].

This paper proposes ASURA, which stands for Advanced Scalable and Uniform storage by Random number Algorithm. In ASURA, data-storing nodes are determined directly by means of a special algorithm that uses pseudorandom numbers (shown as "random numbers" in this paper shortly).

This paper makes two main contributions.
- It proposes ASURA and explains how it can distribute data according to the capacity of each node in any case and can maintain appropriate data distribution with minimum data movement when nodes are added or removed.
- It presents evaluations of ASURA in comparison with Consistent Hashing, Weighted Rendezvous Hashing and Random Slicing, both qualitatively and quantitatively, under equal conditions.

The evaluation results show that:
1. The calculation time of Consistent Hashing, Random Slicing, and ASURA with a thousand nodes was less than 1 micro-second. It was practical for large storage clusters. The calculation time of Weighted Rendezvous Hashing was practical for small storage clusters.
2. Maximum variability in data distribution with Weighted Rendezvous Hashing, Random Slicing and ASURA was far less than 1%. It is better than that of Consistent Hashing. It means storage clusters using Weighted Rendezvous Hashing, Random Slicing and ASURA for distribution algorithm can reduce the number of nodes compared with Consistent Hashing because of the better uniformity of the data distribution.
3. In storage clusters storing data and replication data, Random Slicing needs extra data movements when nodes are added or removed in some cases. It introduces extra states in storage clusters and complexity for programs. Consistent Hashing, Weighted Rendezvous Hashing and ASURA do not need extra data movement when nodes are added or removed.
4. In a real environment, Random Slicing and ASURA achieve both uniform distribution and shorter execution time. Consistent Hashing needs extra nodes for achieving the same capacity because of non-uniformity, and Weighted Rendezvous Hashing needs longer execution time.

Considering the above results, each algorithm has applications suitable for it, and ASURA is better choice for large scale-out storage clusters. ASURA has all characteristics that are shown above; short calculation time, and uniform data distribution, no extra data movement when nodes are added or removed.

The rest of this paper is structured as follows. Section 2 presents related work. Section 3 introduces the algorithm of ASURA. Section 4 discusses qualitative evaluations of the four algorithms, and Section 5 discusses quantitative evaluations of them. Section 6 shows discussion. Section 7 provides a brief summary of the work.

## 2 RELATED WORKS

The author focuses on striping data block algorithms among disks and data distribution algorithms for storages.

RAID [7] is a technique for more reliable storage and more quick accessible storage. It distribute data or calculated data among disks. And it is one of major research area now [8] [9] [10]. At first, it is developed for magnetic disk storages [26] [27] [28]. However it is researched for using in silicon disk storages [29] [30] [31].

Erasure coding is coding techniques by algorithms for reliable data storages. Such a coding technique is researched from the past [32] [33] [34] [35] to today [36] [37] [38].

In early stage of research for data distribution algorithm, algorithms that can distribute data near-uniformly are presented [11] [12]. There is an algorithm that can distribute data with nodes that have non-uniform capacity [13]. Some algorithms achieved data distribution adjustment by parameter adjustment [14] [15]. There are useful algorithms that achieve near-minimum data movement when nodes are added or removed [16] [17]. From now on, this section focuses on algorithms that achieve minimum data movement when nodes are added or removed for clarifying of the contribution of this paper.

"RUSH: Balanced, Decentralized Distribution for Replicated Data in Scalable Storage Clusters" [18] by Honicky and Miller can distribute data uniformly. It achieves minimum data movement when nodes are added or removed. However, it requires the maximum node number in advance. Thus, if the maximum node number is small, it lacks scalability; if the maximum node number is large, it lacks efficient processing.

Karger et al.'s Consistent Hashing in "Consistent Hashing and Random Trees: Distributed Caching Protocols for Relieving Hot Spots on the World Wide Web" [4] is a hash-number-based algorithm for managing data arrangements among a large number of nodes. Nodes are arranged on a ring formed by a number line. A node owns a region extending along the ring from the given node to the nearest node in a given direction. Each datum has its own hash number, which is also located on the ring. The owner of that point on the ring is the data-storing node (Fig. 1). Data distribution is not uniform in the Consistent Hashing. Thus, a virtual node technique has to be applied to achieve uniform distribution. In Consistent Hashing with virtual nodes, each node has many virtual nodes, and the virtual nodes are arranged in many places on the ring of the number line depending on their own hash number. As the number of virtual nodes increases, the sizes of the areas of the nodes approach the same, in accordance with the law of large numbers. Consistent Hashing with virtual nodes achieves both uniform data arrangement and minimum data movement when nodes are added or removed. In addition, Consistent Hashing can be used for distributed storage (P2P system). Even if a node in the Consistent Hashing algorithm knows only some of the nodes, the node can determine a node that is better (i.e., nearer to the right node) than itself. Algorithms [19] [20] [21] have been presented for searching for a data-storing node with information about only some of the node.

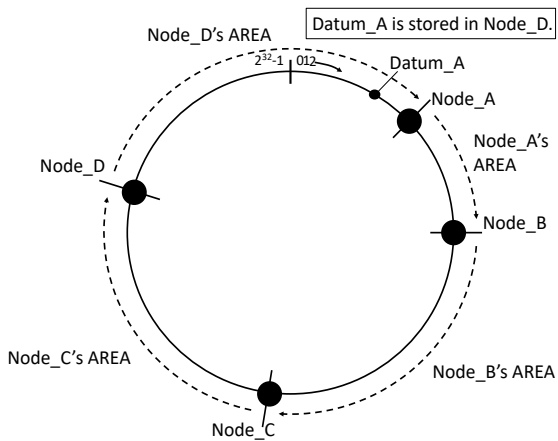

Fig.1 Consistent Hashing

Weil et al.'s Straw Buckets in CRUSH (in "CRUSH: Controlled, Scalable, Decentralized Placement of Replicated Data" [16]) is also an algorithm having preferable characteristics. It is a variety of Highest Random Weight [45]. With it, each node has an individual hash number for an individual datum item, and data are stored in the node having the largest hash number for the data (Fig. 2). Data in this algorithm are distributed uniformly and small number of data movement is required when nodes are added or removed. However it can achieves minimum data movement when nodes are added or removed and uniform data distribution depends on node's capacity in limited case depends on nodes' capacity. CRUSH has four types of algorithms. Each algorithm has advantages and disadvantages.

| Hash numbers of Straw Buckets in CRUSH | | | | | |
|---|---|---|---|---|---|
| | Node_A | Node_B | Node_C | Node_D | Stored Node |
| Data_A | 1 | 5 | (123) | 23 | Node_C |
| Data_B | (99) | 44 | 32 | 56 | Node_A |
| Data_C | 75 | 35 | (76) | 65 | Node_C |
| Data_D | 41 | (97) | 15 | 19 | Node_B |
| Data_E | 11 | 23 | 45 | (68) | Node_D |
| Data_F | 121 | (127) | 112 | 111 | Node_B |

Maximum Hash Number
Hash(Data_F's ID | Node_A's ID)

Fig.2 Straw Buckets in CRUSH

Weighted Rendezvous Hashing [5] is similar algorithm with Straw Buckets in CRUSH. It also give calculated number based on hash number for each node, and data are stored in the node having the largest calculated number for data. Compared with Straw Buckets in CRUSH, it achieves minimum data movement when nodes are added or removed and uniform data distribution depends on node's capacity in any case.

Miranda et al's Random Slicing in "Reliable and Randomized Data Distribution Strategies for Large Scale Storage Systems" [6] also achieves minimum data movements when nodes are added or removed. Random Slicing uses a line having specific length. This line is divided by the area of each node. Then a datum is positioned on this line by hash function and a data-storing node is decided by the position of the datum on this line. When nodes are added, each node's area is shrunk and the new node occupies vacuumed area made by shrinking existing nodes' area. When nodes are removed, removed nodes' areas are deleted, and the remaining node occupies these areas (Fig. 3). In the addition process or the removal process, changing areas are decided so that extra data movements and area fragments are avoided. Random Slicing can distribute data uniformly by equalizing the length of each node, and it has no limitation on scalability.

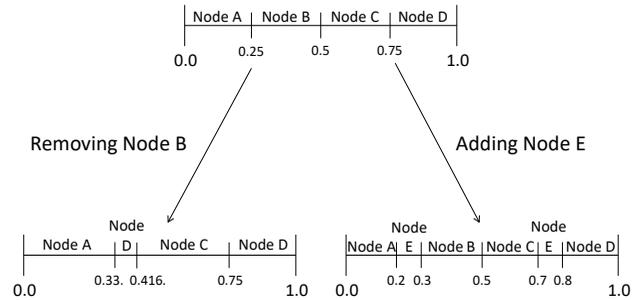

Fig.3 Random Slicing

Chawla et al.'s SPOCA in "Semantics of caching with SPOCA: a stateless, proportional, optimally-consistent addressing algorithm" [22] also achieves minimum data movement when nodes are added or removed. SPOCA assigns segments of a line to nodes. Then the system generates a hash number on this line based on a datum ID until the hash number hits the segment in the line. The node assigned to the segment is the data-storing node. SPOCA can distribute data according to the capacity of each node, and it achieves minimum data movement when nodes are added or removed. However, SPOCA suffers from a trade-off between scalability and efficiency because the length of the line used by SPOCA need to be determined in advance.

Three algorithms (Consistent Hashing, Weighted Rendezvous Hashing, and Random Slicing) achieve minimum data movement to maintain appropriate data distribution when nodes are added or removed, and they also achieve nearly uniform data distribution. Furthermore, they do not impose any limitations on scalability without loss of efficiency in theory. This paper compares them and ASURA both qualitatively and quantitatively.

Data distribution algorithm can use for load balancing.

Researches of load balancing are from unique research [39] [40] [41] to applying data distribution algorithm [42] [43] [44].

## 3 ALGORITHM OF ASURA

This section introduces the algorithm of ASURA.

The precise pseudo code of ASURA is given in Appendix A. It takes as input a datum ID, number of segment, length of each segment, and node number of each segment. Then it outputs the node number of the data-storing node.

Sections 3.1, 3.2, and 3.3 explain the algorithm implemented in this pseudo code. First, the author explains the basic form of ASURA. Second, the author describes how to solve the scalability problem in the basic ASURA by using ASURA random numbers. Third, the author shows how to generate ASURA random numbers. Section 3.4 shows the discussion of node addition and removal in ASURA.

### 3.1 Basic ASURA

This section shows how to achieve appropriate data distribution according to the nodes' capacity, and minimum data movement when nodes are added or removed to keep appropriate data distribution according to the nodes' capacity. The basic algorithm in this section is very similar to SPOCA [22].

The Algorithm of ASURA can be divided into two steps.
  STEP 1.  Assigning nodes to segments in a number line
  STEP 2.  Determining a data-storing node

STEP 1 is the initial step. This step is executed when the system starts and nodes are added or removed. In this step, nodes are assigned to segments in a number line according to the following rules.
1. Nodes are assigned to a segment or segments in the number line. The segment's length is decided by the nodes' capacity. One node can be assigned to multiple segments.
2. The links between the existing nodes and the segments do not change.
3. The segment starts from a point of an integer on the number line. The number of the starting point is a segment number.
4. The segment's length is under 1.0.

Rules 3 and 4 are not necessary, but they can make the program simple and efficient.

Fig. 4 shows an example of nodes' assignments. "X-Y" means greater than or equal to X and less than Y.

The algorithm of STEP 2 is as follows.
1. Initialize a pseudorandom number generator by the datum ID.
2. Generate a random number.
3. If this random number does not point to any segments in the number line, go to 2.
4. If this random number points to a segment in the number line, the node assigned to the pointed segment is determined as the data-storing node.

In this phase, random numbers are generated repeatedly until the random number points to a segment. Theoretically, the random number generation may continue unlimitedly, but it does not occur in the practical usage.

The pseudorandom number generator should have following characteristics (Pseudo Random Number Characteristics).
1. If the seed (the datum ID) is the same, it generates the same random number sequence in a range of pseudo random number.
2. If the seed (the datum ID) is not the same, it generates a different random number sequence in a range of pseudo random number.
3. Random numbers in the random number sequence are nearly homogeneously distributed.

The examples of pseudorandom number generator which has Pseudo Random Number Characteristics are Mersenne Twister [23] [24] and XOR SHIFT [25].

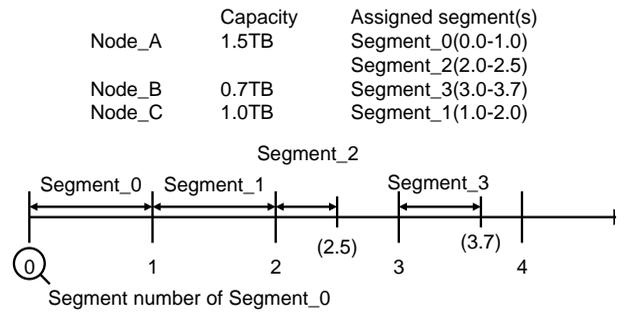

Fig.4 Correspondence of nodes and segments

Fig. 4 shows a sample segment arrange when data-storing nodes for Datum_A and B are determined in the above steps. If random numbers for Datum_A are as follows, the Datum_A is stored in Node_C, which is assigned to Segment_1. Then if random numbers for Datum_B are as follows, the Datum_B is stored in Node_B, which is assigned to Segment_3.
➢ Random numbers for the datum A: 4.2, 1.1…
➢ Random numbers for the datum B: 3.3, 0.6…

In this algorithm, the following characteristics are achieved.
1. When nodes are added, only data that will be stored in added nodes need to be moved. Then data are distributed to nodes according to each node's capacity. The amount of this data movement is minimum.
2. When nodes are removed, only the data that are stored in the removed nodes move. Then data are distributed to nodes according to each node's capacity. The amount of this data movement is minimum.
3. Data are distributed to nodes, nearly according to each node's capacity.

The first characteristic depends on repeatedly generating random numbers until one of them points to a segment. In Fig. 5(a), Node_D, which is assigned to Segment_4, is added to the case in Fig. 4. In this case, if a random number in the random number sequence points to Segment_4 before random numbers in the random number sequence point to Segment_0, 1, 2, and 3 in STEP 2, the datum moves to Node_D, which is assigned to Segment_4. If not, the datum does not move. In this example, Datum_A moves from

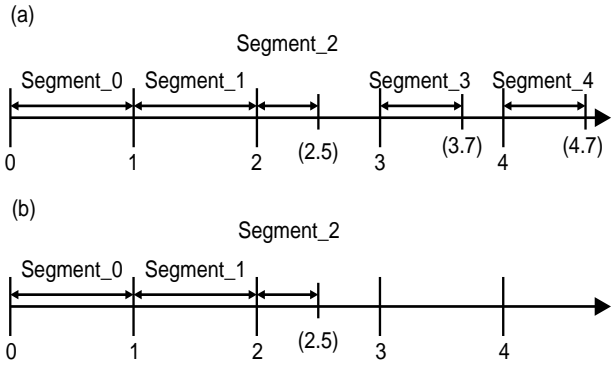
Fig.5 Segment addition and removal by node addition and removal

Node_C to Node_D, and Datum_B does not move. Then data are distributed for nodes according to each node's capacity after Node_D has been added. This means that only minimum amount of data move for keeping the data distribution according to each node's capacity.

The second characteristic also relies on repeatedly generating random numbers until one of them points to a segment. In Fig. 5(b), Node_B, which is assigned to Segment_3, is removed from the case in Fig. 4. In this case, if a random number in the random number sequence pointed to Segment_3 in Fig. 4 before a random number in the random number sequence pointed to Segment_0, 1, and 2 in STEP 2, the datum moves to another node. If not, the datum does not move. In this example, Datum_A does not move, and Datum_B moves from Node_B to Node_A. Then data are distributed to nodes according to each node's capacity after Node_B has been removed. This means that only minimum amount of data move from removed nodes in order to keep data distribution according to each node's capacity.

The third characteristic is obvious, because the segment length of each node is decided by the node's capacity.

However, the basic ASURA lacks scalability or efficiency. If the maximum random number is small, ASURA lacks scalability, because after the number line is filled with segments, no nodes can be added; if the maximum random number is large, it lacks efficiency, because a lot of random numbers should be generated until it points to a segment. The next section explains how to solve this problem.

### 3.2 ASURA random number

This section describes ASURA random numbers, which are used in ASURA. ASURA random numbers can extend or shrink its range using a special method. It makes ASURA both scalable and efficient by expanding or shrinking the number line.

The random number sequence for ASURA needs to have Pseudo Random Number Characteristics. Thus, ASURA random numbers also need to have the same characteristics.

At first, this paper shows how to extend the random number range.

In the example given in Fig. 4, a random number generator that generates 0.0-4.0 can be used. An example of the random number sequence is as follows.
- Random number sequence: 2.7, 3.8, 1.1 …

In this random number sequence, Node_C is the data-storing node.

If other random numbers, which are 4.0–8.0 and which are generated from a certain seed based on the datum ID, are inserted into this random number sequence while keeping the same generation probability, this example becomes as follows.
- Number sequence: <u>5.4</u>, 2.7, <u>4.3</u>, 3.8, <u>7.2</u>, 1.1 … (Underlined numbers are inserted)

In this number sequence, random numbers that are 0.0–4.0 keep their values and order. Thus, the data-storing node is not changed from Node_C even after the insertion of random numbers that are 4.0–8.0 in the case of Fig. 4.

Generally, the selection of the data-storing node assigned to segments in the original random number area is unchanged.

Then the whole random number sequence has the Pseudo Random Number Characteristics. It means that they can be used as ASURA random numbers. The range of random number sequences can be extended without side effects by inserting random numbers whose range is outside its number range.

Shrinking the range of random numbers can be discussed in the same way. The range of random number sequences can be shrunk unlimitedly by removing numbers that are outside the new range, which covers the assignment area of segments.

In conclusion, the range of random number sequences used by ASURA can be extended or shrunk. By covering a wider area, the range of ASURA's random numbers can be extended for scalability without any side effects. If segments are changed to cover a narrower area, the range can be shrunk for efficiency without any side effects. Overall, ASURA can achieve both scalability and efficiency by using ASURA random numbers. The next section describes how to generate ASURA random numbers.

### 3.3 Generation of ASURA random numbers

The algorithm for generating ASURA random numbers is simple.

The algorithm for generating ASURA random numbers uses several base pseudorandom number generators. Each base pseudorandom number generator need to have the following specifications.
1. It generates a different range of random numbers.
2. The range of random numbers generated by a base pseudorandom number generator that has a wider range of random numbers must cover the range of random numbers generated by one that has a narrower range of random numbers.
3. It has Pseudo Random Number Characteristics.

An example of the base pseudorandom number generator is Mersenne Twister [24] and XOR SHIFT [25].

The base pseudorandom number generator can generate random numbers using a seed that can be calculated from a datum ID.

Then ASURA random numbers are generated by the following algorithm.
1. A base pseudorandom number generator having the narrowest range of random numbers that cover all of the segments is selected.
2. The selected base pseudorandom number generator generates a random number.

3. If this random number is in the range of the base pseudorandom number generator having the one-step narrower range, the one-step narrower range of base random number generator is selected, and the algorithm goes to 2.
4. If this random number is not in the range of the base pseudorandom number generator having the one-step narrower range, or if there are no narrower range of base random number generator, the random number is an ASURA random number.

The author will explain the algorithm using the following example.
1. The first range of ASURA random numbers is 0.0–4.0.
2. The extended range of ASURA random numbers is 0.0–8.0.
3. Twice the extended range of ASURA random numbers is 0.0–16.0.

Three base pseudorandom number generators are used in these cases. The first (Generator-0) generates random numbers that are 0.0-4.0. The second (Generator-1) generates random numbers that are 0.0-8.0. The third (Generator-2) generates random numbers that are 0.0-16.0

First, let us consider the generation of ASURA random numbers that are 0.0–4.0. In this case, the Generator-0 is used simply. For example,
➢ ASURA random numbers: 2.7, 3.8, 1.1 …

Second, let us consider the generation of ASURA random numbers that are 0.0–8.0. In this case, Generator-0 and Generator-1 are used. First, Generator-1, which has the narrowest range of random numbers covering all of the segments, generates a random number. If the generated random number is 4.0–8.0, it is an ASURA random number, and if it is 0.0–4.0, then a random number generated by Generator-0, which has the one-step narrower range of random numbers, is an ASURA random number. For example,
➢ Random numbers generated by Generator-0: 2.7, 3.8, 1.1 …
➢ Random numbers generated by Generator-1: 5.4, 3.2, 4.3, 2.2, 7.2 …
➢ ASURA random numbers: 5.4, 2.7, 4.3, 3.8, 7.2 … (Underlined numbers come from Generator-0)

Comparing ASURA random numbers that are 0.0–4.0 and ones that are 0.0–8.0, the latter includes the former by keeping its value and order; random numbers that are 4.0–8.0 are inserted in the ASURA random numbers that are 0.0–4.0. Random numbers in the ASURA random numbers that are 0.0–8.0 are nearly homogeneously distributed, because the random numbers that are 0.0–4.0, which were generated by the Generator-1, are replaced with those that were generated by Generator-0. This meets the requirements of ASURA random numbers.

The generation of ASURA random numbers that are 0.0–16.0 is similar. It uses the Generator-0, Generator-1, and Generator-2.
➢ Random numbers generated by Generator-0: 2.7, 3.8, 1.1 …
➢ Random numbers generated by Generator-1: 5.4, 3.2, 4.3, 2.2, 7.2 …
➢ Random numbers generated by Generator-2: 9.2, 2.8, 8.0, 4.4, 9.1, 1.2 …
➢ ASURA random numbers: 9.2, 5.4, 8.0, 2.7, 9.1, 4.3 … (An underlined number comes from Generator-0; wave-lined numbers come from Generator-1.)

Obviously, this meets the requirements of ASURA random numbers.

ASURA random numbers can be extended without limit in the same manner.

When ASURA random numbers are shrunk, only unnecessary base pseudorandom number generators are eliminated. The result of eliminating pseudorandom number generators is obvious.

The pseudo code in Appendix A presents STEP 2 in ASURA discussed from section 3.1 to section 3.3.

This paper applies ASURA to a limited one-dimensional line, but ASURA can be applied to general one-dimensional lines or even multidimensional space. However, this paper does not discuss them because they are not the scope of this paper.

### 3.4 Node addition and removal

This section explains discussion of solving problems of node addition and node removal.

First, all nodes that execute ASURA must have the same pairs of nodes and segments. Thus, coordination of nodes and segments must be done in a centralized manner when the storage cluster is initialized or when nodes are added or removed. This can be done by a temporary central node, and every node can be the temporary central node. Thus, the temporary central node does not become SPoF.

Second, recalculating the data-storing node is a costly process. However, its cost is small compared with that of transferring data to another node through the network. Thus, the cost of recalculating the data-storing node does not become a problem. Then recalculation of the data-storing node can be executed on all nodes in parallel. Thus, even if the number of nodes is many, the scalability problem does not occur.

## 4 QUALITATIVE EVALUATION

This section discusses qualitative evaluations of Consistent Hashing, Weighted Rendezvous Hashing, Random Slicing, and ASURA.

In the following evaluations, the author focuses on algorithms that achieve minimum data movements for keeping an appropriate data distribution when nodes are added or removed and achieve both efficiency and scalability theoretically; algorithms that do not meet these requirements, for example Rush_R, Straw Buckets in CRUSH and SPOCA, were not evaluated. It makes the comparison standards clear and makes the comparisons easy.

In following chapter, Consistent Hashing is called for CH, Weighted Rendezvous Hashing is called for WRH, Random Slicing is called RS and Virtual nodes is called VN if needed.

The evaluations in this chapter focus on scalability because the algorithms are to be used for data management in large scale-out storage clusters.

Distribution algorithms are evaluated in terms of three factors which appear difference between algorithms: calculation time, distribution uniformity and data movement with replication.

The following parameters are used:
  N: number of nodes
  V: number of virtual nodes in each node

### 4.1 Calculation time

Consistent Hashing calculates the hash number of a datum in the distribution stage from the datum ID and searches for the relevant data-storing node using a binary search, for example. Thus, the order of the calculation time in the distribution stage is O (log (NV)). Karger et al. presented an algorithm whose order of calculation time in the distribution stage is O (1) [4].

Weighted Rendezvous Hashing calculates the hash numbers of nodes from data IDs and node IDs, and it searches for the largest hash number from the calculated numbers which are calculated from these hash numbers. Thus, the order of calculation time in Weighted Rendezvous Hashing is O (N).

Random Slicing can be treated as the same as Consistent Hashing. Thus the order of the calculation time in the distribution stage is O (log (NX)). X becomes larger when N becomes larger. Then X is depend on each node's capacity, thus it can't define by equation. O (1) algorithm can be used for Random Slicing as same as Consistent Hashing.

ASURA generates random numbers from the seeds until it points to a segment. As the number of nodes increases, number of repetitions by the pseudorandom number generator approaches a constant value. For generating ASURA random number, the probability of using the first base random number is 1, that of second one is 1/2, that of third one is 1/4, etc. Therefore the expectation number of repetitions becomes 2. And the repetition number of generating ASURA random number depends on the ratio of the length of the number line that does not have a corresponding segment (the proof of this is given in Appendix B), but does not depends on the length of the number line that does not have a corresponding segment. Thus, the maximum order of calculation time of ASURA is O (1).

With respect to calculation times, Weighted Rendezvous Hashing is the worst in terms of scalability, Consistent Hashing is the second best or the best, and Random Slicing is also the second best or the best and ASURA is the best.

### 4.2 Distribution uniformity

In Consistent Hashing, the hash numbers of virtual nodes have variability, and those of data also have variability, which means that distribution of Consistent Hashing suffers from double variability. In Weighted Rendezvous Hashing and in Random Slicing, hash numbers have variability. In ASURA, ASURA random numbers have variability. Therefore, these three suffer from only single variability. With respect to the distribution uniformity, Weighted Rendezvous Hashing, Random Slicing, and ASURA are better than Consistent Hashing.

### 4.3 Data movement with replication

In large scale-out storage clusters, node failure is usual because it composes many nodes. Thus, data should be replicated to some nodes for failure recovery.

All the four algorithms can decide replication nodes. For example, Weighted Rendezvous Hashing can decide the first replication node by selecting a node having the second largest hash number. Consistent Hashing can decide the first replication node by selecting the next node in the given direction. Random Slicing can decide the first replication nodes by re-generating hash number. And ASURA can decide the first replication nodes by repeating generating random numbers until the number is in an existing segment. Consistent Hashing, Random Slicing, and ASURA need to check the selected first replication node is different from the data-storing node.

When one node is added to a storage cluster, in the case of Consistent Hashing, Weighted Rendezvous Hashing, and ASURA, no datum or only one datum per one datum moves from a data-storing node or replication nodes to the new node.

However, in the case of Random Slicing, several data per one datum may move from a data-storing node or replication nodes to the other node. For example, please consider a case that the replication of datum is in area-A, and the data-storing node of the datum is changed to the added node. Here, there might be a case that the whole part of the area-A is changed to the added node's area. In this case, since both the original datum and the replication is in the same node, the replication need to be moved to another node, which requires extra data movement compared to the other three algorithms.

In a storage cluster using replication, Consistent Hashing, Weighted Rendezvous Hashing, and ASURA are better than Random Slicing in terms of the amount of data movement.

### 4.4 Summary of qualitative evaluation

The results of the qualitative evaluation are summarized in Table I. Each algorithm was found to have its own advantages and disadvantages.

## 5. QUANTITATIVE EVALUATION

This section discusses quantitative evaluations of Consistent Hashing, Weighted Rendezvous Hashing, Random Slicing, and ASURA.

It is assumed that capacity of the nodes is fixed in most evaluations, because flexible capacity would invite too much complexity to comparison.

Evaluation conditions are as follows:

Table I. Qualitative Evaluation Results

|  | Calculation time | | Distribution uniformity | | Data movement with replication | |
|---|---|---|---|---|---|---|
| CH | **Medium or Good** | O(log(NV)) or O(1) | **Poor** | Double variability | **Good** | No extra data movement |
| WRH | **Poor** | O(N) | **Good** | Single variability | **Good** | No extra data movement |
| RS | **Medium or Good** | O(log(NX)) or O(1) | **Good** | Single variability | **Poor** | Need extra data movement |
| ASURA | **Good** | O(1) | **Good** | Single variability | **Good** | No extra data movement |

| CPU | Intel Xeon E5504 2.0GHz |
| --- | --- |
| Memory | 12GB |
| OS | Ubuntu 12.10(Kernel 3.5.0-26-generic) |
| Gcc | gcc 4.7.2 |
| SFMT | dSFMT 2.1 compiled with -O6 -march=native -DDSFMT_MEXP=521 |

### 5.1 Calculation time

This evaluation focuses on the practicality of calculation time and its scalability. Small differences in calculation time are not significant, because the algorithms are to be used with network access.

The following conditions are assumed:

For all four algorithms —

The number of nodes range from 1 to 1000. The calculation time is evaluated on the basis of execution time of 1,000,000 loops for different inputs. The evaluation is done 100 times. And the result is average of 60 results in the center.

For Consistent Hashing —

The numbers of nodes tested are 1, 50, 100 … 950, 1000. The numbers of virtual nodes tested are 100, 1000 and 10,000. In the initial stage, hash numbers of virtual nodes are random numbers generated by SFMT, and the hash numbers of virtual nodes are sorted. Data-storing nodes are searched for in a table in memory with binary search.

For WRH —

Hash numbers for each node are generated by SFMT and are compared on the fly with the current maximum number.

For RS —

The numbers of nodes tested are 1, 50, 100 … 950, 1000. Hash numbers are generated by SFMT and compared with a binary search.

For ASURA —

The random numbers output by the first pseudorandom number generator are 0.0–16.0. The maximum numbers of random numbers output by the following base pseudorandom number generators are doubled when the number of nodes is increased. The nodes are numbered consecutively, with no gaps between segment numbers.

The evaluation results tests are plotted in Fig. 6. The calculation times of Consistent Hashing, Random Slicing, and ASURA are short enough to be practical when data transfer in network communications are taken into consideration. The calculation time of Weighted Rendezvous Hashing is over 0.50 micro-seconds in 12 nodes and it extends linearly beyond the graph area. Evaluation of O (1) algorithm of Consistent Hashing is skipped. Because O (log (NV)) algorithm of Consistent Hashing is fast enough and it is clear that O (1) has problems of memory consumption or complex treatment of hash number. And O (1) algorithm of Consistent Hashing is not used in real in author's best knowledge. O (1) algorithm of Random Slicing is not evaluated because of the same reason. The calculation time of ASURA is increase dramatically when base pseudo random number generator is added as number of node is increase. Thus the graph of ASURA's calculation time looks like saw tooth. This evaluation demonstrates that Consistent Hashing, Random Slicing, and ASURA are realistic choices for large storage clusters and that Weighted Rendezvous Hashing suits for small storage clusters.

### 5.2 Distribution uniformity

Distribution uniformity in Consistent Hashing depends on the number of virtual nodes, which is varied in the distribution uniformity tests. With Weighted Rendezvous Hashing, Random Slicing and ASURA, distribution uniformity depends on the number of data per node, and this number is also varied in the distribution uniformity tests.

Conditions of the distribution uniformity test are as follows:

Number of nodes: 10 100 and 1000

Number of virtual nodes in each node (for Consistent Hashing only): 100, 1000 and 10,000

Number of data in each node: 1000, 3162, 10,000, 31,622 and 100,000

The test is looped 20 times.

The results are shown in Figs. 7–9. The figures indicate that Consistent Hashing has lower uniformity. Weighted Rendezvous Hashing, Random Slicing, and ASURA achieve the same level of uniformity. Weighted Rendezvous Hashing, Random Slicing and ASURA are able to achieve less than 1% maximum variability in the best case, while Consistent Hashing is able to achieve only a few percent level maximum variability in the best case.

Distribution uniformity with non-uniform capacity nodes is also evaluated. The test conditions are as follows:

Number of nodes: 100

Number of expected storing data: 1,000,000 × node number (starting from 1)

Number of virtual nodes for each node (for Consistent Hashing only): {1, 10, 100} × node number

Number of data: 5,050,000,000

Number of tests: 20 times.

The results in Table II reveal that Weighted Rendezvous Hashing, Random Slicing and ASURA will have better uniformity than Consistent Hashing in storage clusters with nodes having non-uniform capacity.

### 5.3 Redistribution

Evaluating data redistribution when nodes are added or removed is difficult, because it is affected by many factors. For example,

1. Number of data in one node

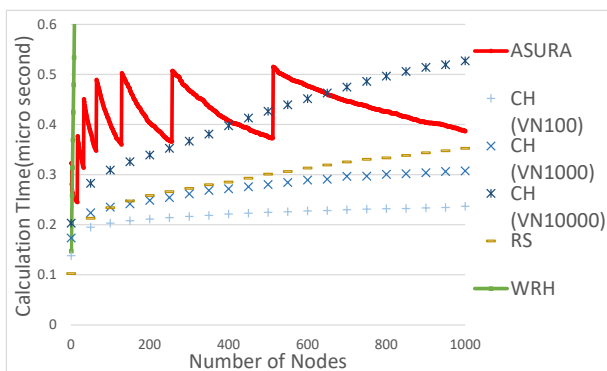

Fig.6 Calculation time

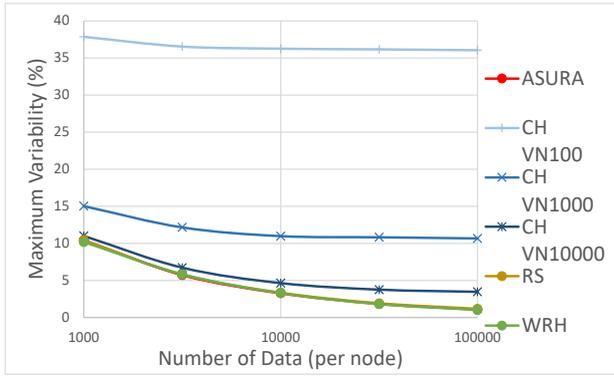
Fig.7 Maximum variability (10 nodes)

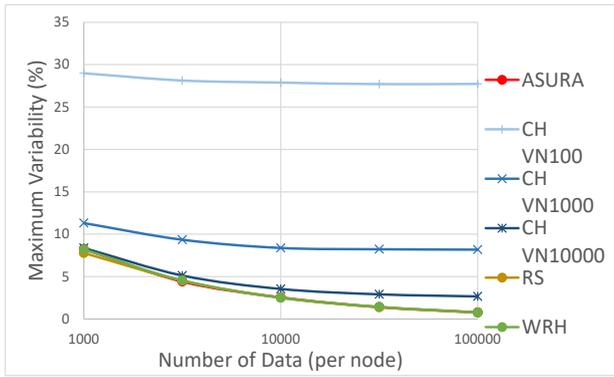
Fig.8 Maximum variability (100 nodes)

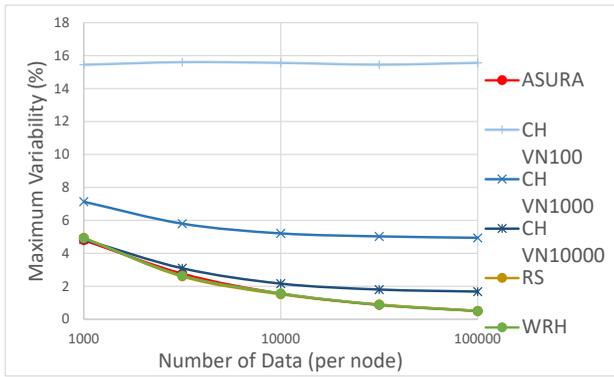
Fig.9 Maximum variability (1000 nodes)

2. Number of data that are transferred to new data-storing node
3. Calculation time of one data-storing node
4. Accessing time of datum
5. Transferring time of datum

In the above factors, 4 and 5 depends greatly on environments. Thus this evaluation focuses on 1, 2, and 3 that are less dependent on environments.

The evaluation conditions are as follows:
  Number of nodes: changed from 16 to 17
  Number of storing data in storage clusters: 16,000,000
  Number of virtual nodes for each node (for Consistent Hashing only): 100, 1000, 10000
  Number of tests: 20 times

The results are shown in Table III, which reveal that Random Slicing and ASURA have better characteristics and all algorithms can be practically used in the case of redistribution.

Table II. Variability of non-uniform capacity nodes (Decorated value is better)

|  | Maximum number of data compared with the theoretical number | Minimum number of data compared with the theoretical number |
|---|---|---|
| CH with 100 maximum VN | +34.2% - +372.8% | -32.6% - -83.9% |
| CH with 1000 maximum VN | +11.1% - +78.1% | -10.8% - -44.2% |
| CH with 10000 maximum VN | +4.4% - +34.9% | -3.9% - -13.8% |
| WRH | ***+0.04% - +0.24%*** | ***-0.06% - -0.31%*** |
| RS | ***+0.05% - +0.14%*** | ***-0.07% - -0.28%*** |
| ASURA | ***+0.06% - +0.09%*** | ***-0.07% - -0.09%*** |

Table. III Redistribution cost (Decorated value is better)

|  | Maximum number of data in one node | Calculation time of one data – storing node (microsecond) | Maximum number of transferring data |
|---|---|---|---|
| CH (VN 100) | 1184080.30 | ***0.1808*** | 136229.15 |
| CH (VN 1000) | 1060850.85 | ***0.2077*** | 80306.65 |
| CH (VN 10000) | 1017349.75 | ***0.2442*** | 65197.65 |
| WRH | ***1001951.00*** | 0.9829 | ***59296.10*** |
| RS | ***1001826.40*** | 0.1747 | ***59210.15*** |
| ASURA | ***1001756.05*** | 0.2457 | ***59242.75*** |

### 5.4 Redistribution with replication

Redistribution with replication is tested. In this test, it is evaluated that number of moving data between nodes when a node is added and a node is removed. Data are replicated by 3, storing data, 1st replicated data, and 2nd replicated data. Number of node becomes 9 from 8 and 8 from 9. Number of master data is 1,000,000. This test is looped in three times. The results are shown in Table IV. –A means a case of node addition and –R means a case of node removal. 0 shows number of data which don't move any replicated data when a node is added or removed. 1 shows number of data which move 1 datum in storing datum and 2 replicated data when node is added or removed. 2 and 3 shows in the same way. This results reveal that Consistent

Table. IV Redistribution with Replication (Decorated value is better)

|  | 0 | 1 | 2 | 3 |
|---|---|---|---|---|
| CH-A | 66553833 | 33446167 | ***0*** | ***0*** |
| CH-R | 66554519 | 33445481 | ***0*** | ***0*** |
| WRH-A | 66666890 | 33333110 | ***0*** | ***0*** |
| WRH-R | 66669256 | 33330744 | ***0*** | ***0*** |
| RS-A | 66667393 | 31397312 | 1901370 | 33925 |
| RS-R | 66680268 | 31387070 | 1899271 | 33391 |
| ASURA-A | 66661333 | 33338667 | ***0*** | ***0*** |
| ASURA-R | 66669190 | 33330810 | ***0*** | ***0*** |

Hashing, Weighted Rendezvous Hashing, ASURA needs only minimum data movements when a node is added and a node is removed. However Random Slicing needs extra data movements in the same case.

### 5.5 Real Environment

For real environment test, memcached, memslap in libmemcached and modified twemproxy are used.

The test environments are

Memcached server:
 CPU: INTEL Xeon E5-2620 v4 2.10GHz
 Memory: 512GB
 Network: 1000BASE-T
 OS: Ubuntu 16.04(Kernel 4.4.0-57-generic)
 Gcc: 5.4.0

Memslap and twemproxy server:
 CPU: INTEL Xeon E5-2620 v4 2.10GHz
 Memory: 256GB
 Network: 1000BASE-T
 OS: CentOS 6.3(Kernel 2.6.32-642.11.1.el6.x86_64)
 Gcc: 4.4.7

Memcached 1.4.4-3.el6:
 Option -d
 Number of server 1
 Number of memcached 100

Memslap in libmencached 0.31-1.1.el6:
 Option --execute-number=1,000,000

Modified twemproxy 0.2.2:
 Compile option –O6 –march=native –mfpmath=sse –msse4.2
 Modified for using Weighted Rendezvous Hashing, Random Slicing, and ASURA
 Consistent Hashing: 160 virtual nodes
 ASURA: Initial maximum random number is 16

Data are set from memslap to memcached server through twemproxy server 1,000,000 times. This test is looped 10 times.

The evaluation results are shown in Table V.

Table V. Real environment (Decorated value is better)

|        | Average Execution Time | Maximum Variability |
|--------|------------------------|---------------------|
| CH     | ***232.5s***           | 19.46%              |
| WRH    | 285.3s                 | ***0.89%***         |
| RS     | ***235.5s***           | ***0.94%***         |
| ASURA  | ***236.0s***           | ***1.20%***         |

The results show that Random Slicing and ASURA show better characteristics than Consistent Hashing and Weighted Rendezvous Hashing. Consistent Hashing has large maximum variability, which leads less usage of storage capacity. Weighted Rendezvous Hashing takes larger execution time.

### 5.6 Summary of quantitative evaluations

At first, these evaluations show that ASURA has 4 basic characteristics: 1) minimum data movement to maintain data distribution according to node capacity when nodes are added or removed, even if data are replicated, 2) roughly sub-micro-seconds calculation time, 3) much lower than 1% maximum variability between nodes in data distribution, and 4) data distribution according to the capacity of each node.

The calculation time for Consistent Hashing, Random Slicing, and ASURA is very short (under 1 micro-second). Therefore, these three can be used for large storage clusters. The calculation time of Weighted Rendezvous Hashing is longer than the other three clearly. Thus Weighted Rendezvous Hashing suits for small storage clusters.

The variability of data distribution between nodes of Weighted Rendezvous Hashing, Random Slicing and ASURA is better than Consistent Hashing. If variability of data distribution between nodes is high, storage cluster cannot fully utilize its capacity, because one node becomes full of data when the other nodes still have enough vacancy.

When data are replicated, the number of moving data between nodes when a node is added or removed in storage using for Random Slicing is larger than storage using for Consistent Hashing, Weighted Rendezvous Hashing or ASURA. This extra data movement makes process complicate and slow.

From all results in this paper, four algorithms have suitable applications for each. Consistent Hashing is the only choice for distributed storages (P2P system). Weighted Rendezvous Hashing meets small storage clusters. Random Slicing is applicable for both small and large storage clusters without data replication. Then ASURA fits for both small and large storage clusters with data replication. It means that ASURA is suitable for large scale-out storage clusters.

## 6 DISCUSSION

### 6.1 Meaning of uniform distribution

Uniform data distribution achieves node reduction in storage system.

In usual distribution storage using algorithm distribution, the storage stores data to a node which is indicated by the algorithm. And if the node which is indicated by the algorithm is full, the storage can't store data more. If the storage stores the data to another node, the storage becomes very complex and very inefficient. Then when data are added in the storage using un-uniform data distribution algorithm, one node becomes full before other nodes have enough data. These mean that the storage using un-uniform data distribution algorithm can't use storage capacity fully and needs more node for having the same capacity, compared with the storage using uniform data distribution algorithm.

If distribution algorithm has 10% maximum variability, a storage using this distribution algorithm needs 9.1% times the number of nodes additionally than storage using ideal distribution algorithm.

A storage using Weighted Rendezvous Hashing, Random Slicing and ASURA, which achieve uniform distribution, can reduce nodes dramatically than a storage using Consistent Hashing, which can't achieve uniform distribution.

### 6.2 Uniform data distribution and various size of data and various access frequency of data

Uniform data distribution is meaningful even if data has various sizes or various access frequencies. If data distribution in the storage is uniform, the storage suffers single un-uniformity; data size or access frequency of data. However if data distribution in the storage is un-uniform, the storage suffers double un-uniformity; data distribution and data size or access frequency of data. It causes more un-uniform usage of data storing capacity and un-uniform data access. It means that uniform data distribution has meaning for storage storing data having various sizes or data having various access frequencies.

# 7 CONCLUSION

This paper presented ASURA, which can distribute data according to node capacity and can retain uniformity with minimum data movements when nodes are added or removed. It needs only limited resources for its execution and has scalability. This paper contrasted it with similar algorithms, i.e., Consistent Hashing, Weighted Rendezvous Hashing, and Random Slicing. The evaluation results show that each algorithm has suitable use cases. Then they show that ASURA meets data distribution algorithm for large scale-out storage clusters.

## REFERENCES

[1] S. Ghemawat, H. Gobioff and S. Leung, "The Google file system," In Proceedings of the nineteenth ACM symposium on Operating systems principles, October 2003, New York, USA, 29-43.

[2] G. Decandia, D. Hastorun, M. Jampani, G. Kakulapati, A. Lakshman, A. Pilchin, S. Sivasubramanian, P. Vosshall and W. Vogels, "Dynamo: amazon's highly available key-value store," In Proceedings of twenty-first ACM SIGOPS symposium on Operating systems principles, New York, USA, October 2007, 205-220.

[3] S. A. Weil, S. A. Brandt, E. L. Miller, D. D. E. Long and C. Maltzahn, "Ceph: a scalable, high-performance distributed file system," OSDI '06 Proceedings of the 7th symposium on Operating systems design and implementation, California, USA, November 2006, 307-320.

[4] D.R. Karger, E. Lehman, T. Leighton, R. Panigrahy, M. Levine and D. Lewin, "Consistent hashing and random trees: distributed caching protocols for relieving hot spots on the World Wide Web," Proceedings of the twenty-ninth annual ACM symposium on Theory of computing, Texas, USA, May 1997, 654–663.

[5] http://www.snia.org/sites/default/files/SDC15_presentations/dist_sys/Jason_Resch_New_Consistent_Hashings_Rev.pdf

[6] A. Miranda, S. Effert, Y. Kang, E. L. Miller, A. Brickmann and T. Cortes "Reliable and Randomized Data Distribution Strategies for Large Scale Storage Systems," in Proceedings of the 18th International Conference on High Performance Computing (HiPC), 2011.

[7] Patterson, David A., Garth Gibson, and Randy H. Katz. A case for redundant arrays of inexpensive disks (RAID). Vol. 17. No. 3. ACM, 1988.

[8] Lin, Sian-Jheng, Amira Alloum, and Tareq Y. Al-Naffouri. "RAID-6 reed-solomon codes with asymptotically optimal arithmetic complexities." Personal, Indoor, and Mobile Radio Communications (PIMRC), 2016 IEEE 27th Annual International Symposium on. IEEE, 2016.

[9] Wang, Mingyang, and Yiming Hu. "i-RAID: a novel redundant storage architecture for improving reliability, performance, and life-span of solid-state disk systems." Proceedings of the 31st Annual ACM Symposium on Applied Computing. ACM, 2016.

[10] Fu, Yingxun, et al. "Short Code: An Efficient RAID-6 MDS Code for Optimizing Degraded Reads and Partial Stripe Writes." IEEE Transactions on Computers 66.1 (2017): 127-137.

[11] R. Devine, "Design and Implementation of DDH: A Distributed Dynamic Hashing Algorithm," Foundations of Data Organization and Algorithms, 730, 101–114, 1993.

[12] W. Litwin, M. Neimat and D.A. Schneider, "LH*-a scalable, distributed data structure," ACM Transactions on Database Systems, 21 4, 480–525, 1996.

[13] A. Brinkmann, K. Salzwedel and C. Schcideler "Compact, adaptive placement schemes for non-uniform distribution requirements," In proceedings of the 14th ACM Symposium on Parallel Algorithms and Architectures (SPAA), pages 53-62, Winnipeg, Manitoba, Canada, 2002.

[14] C. Wu and R. Burns, "Handling Heterogeneity in Shared-Disk File Systems," Proceedings of the 2003 ACM/IEEE conference on Supercomputing, Arizona, USA, November 2003, 7–19.

[15] C. Wu and R. Burns, "Tunable randomization for load management in shared-disk clusters." Trans. Storage 1, 1, 108–131, 2005.

[16] S.A. Weil, S.A. Brandt, E.L.Miller and C. Maltzahn, "CRUSH: controlled, scalable, decentralized placement of replicated data," Proceedings of the 2006 ACM/IEEE conference on Supercomputing, Florida, USA, Article 122.

[17] M. Mense and C. Scheideler, "SPREAD: an adaptive scheme for redundant and fair storage in dynamic heterogeneous storage systems," In Proceedings of the nineteenth annual ACM-SIAM symposium on Discrete algorithms, Philadelphia, USA, January 2008, 11

[18] R. J. Honicky and E. L. Miller, "Replication under scalable hashing: A family of algorithms for scalable decentralized data distribution" In Proceedings of the 18th International Parallel & Distributed Processing Symposium, New Mexico, USA, April, 2004, 96.

[19] I. Stoica, R. Morris, D.R. Karger, M.F. Kaashoek and H. Balakrishnan, "Chord: A scalable peer-to-peer lookup service for internet applications," Proceedings of the international conference on Applications, technologies, architectures, and protocols for computer communications, California, USA, August 2001, 149–160.

[20] T. Rowstron and P. Druschel, "Pastry: Scalable, Decentralized Object Location, and Routing for Large-Scale Peer-to-Peer Systems," Proceedings of the IFIP/ACM International Conference on Distributed Systems Platforms, Heidelberg, Germany, November 2001, 329-350.

[21] B.Y. Zhao, J.D. Kubiatowicz and A.D. Joseph, "Tapestry: An Infrastructure for Fault-tolerant Wide-area Location and Routing," University of California at Berkeley, Berkeley, CA, 2001.

[22] A. Chawla, B. Reed, K Juhnke and G Syed, "Semantics of caching with SPOCA: a stateless, proportional, optimally-consistent addressing algorithm," Proceedings of the 2011 USENIX conference on USENIX annual technical conference, California, USA, June 2011, 33.

[23] M. Matsumoto and T. Nichimura, "Mersenne twister: a 623-dimensionally equidistributed uniform pseudo-random number generator," ACM Transactions on Modeling and Computer Simulation, 8, 1, 3–30, 1998.

[24] M. Saito and M. Matsumoto, "SIMD-oriented fast mersenne twister: a 128-bit pseudorandom number generator," In Monte-Carlo and Quasi-Monte Carlo Methods, 607–622, 2006.

[25] Marsaglia, George. "Xorshift rngs." Journal of Statistical Software 8.14 (2003): 1-6.

[26] Peter M. Chen and David A. Patterson. 1990. Maximizing performance in a striped disk array. In Proceedings of the 17th annual international symposium on Computer Architecture (ISCA '90). ACM, New York, NY, USA, 322-331. DOI=http://dx.doi.org/10.1145/325164.325158

[27] Gray, Jim, Bob Horst, and Mark Walker. "Parity Striping of Disk Arrays:

Low-Cost Reliable Storage with Acceptable Throughput." VLDB. 1990.

[28] Im, Soojun, and Dongkun Shin. "Flash-aware RAID techniques for dependable and high-performance flash memory SSD." IEEE Transactions on Computers 60.1 (2011): 80-92.

[29] Lee, Sehwan, et al. "A lifespan-aware reliability scheme for RAID-based flash storage." Proceedings of the 2011 ACM Symposium on Applied Computing. ACM, 2011.

[30] Lee, Yangsup, Sanghyuk Jung, and Yong Ho Song. "FRA: a flash-aware redundancy array of flash storage devices." Proceedings of the 7th IEEE/ACM international conference on Hardware/software codesign and system synthesis. ACM, 2009.

[31] Blaum, Mario, James Lee Hafner, and Steven Hetzler. "Partial-MDS codes and their application to RAID type of architectures." IEEE Transactions on Information Theory 59.7 (2013): 4510-4519.

[32] Peterson, William Wesley, and Edward J. Weldon. Error-correcting codes. MIT press, 1972.

[33] MacWilliams, Florence Jessie, and Neil James Alexander Sloane. The theory of error-correcting codes. Elsevier, 1977.

[34] Blahut, Richard E. Theory and practice of error control codes. Vol. 126. Reading (Ma) etc.: Addison-Wesley, 1983.

[35] S.Lin, D.J.Costello, "Error Control Coding: Fundamentals and Applications", Prentice Hall, 1983.

[36] Li, Runhui, Yuchong Hu, and Patrick PC Lee. "Enabling efficient and reliable transition from replication to erasure coding for clustered file systems." IEEE Transactions on Parallel and Distributed Systems (2017).

[37] Zhang, Heng, Mingkai Dong, and Haibo Chen. "Efficient and Available In-memory KV-Store with Hybrid Erasure Coding and Replication." FAST. 2016.

[38] Burihabwa, Dorian, et al. "A performance evaluation of erasure coding libraries for cloud-based data stores." Distributed Applications and Interoperable Systems. Springer International Publishing, 2016.

[39] Azar, Yossi, et al. "Balanced allocations." SIAM journal on computing 29.1 (1999): 180-200.

[40] Pagh, Rasmus, and Flemming Friche Rodler. "Cuckoo hashing." Journal of Algorithms 51.2 (2004): 122-144.

[41] Talwar, Kunal, and Udi Wieder. "Balanced allocations: A simple proof for the heavily loaded case." International Colloquium on Automata, Languages, and Programming. Springer Berlin Heidelberg, 2014.

[42] Byers, John, Jeffrey Considine ^⋆, and Michael Mitzenmacher. "Simple load balancing for distributed hash tables." Peer-to-peer systems II (2003): 80-87.

[43] Byers, John W., Jeffrey Considine, and Michael Mitzenmacher. "Geometric generalizations of the power of two choices." Proceedings of the sixteenth annual ACM symposium on Parallelism in algorithms and architectures. ACM, 2004.

[44] Mirrokni, Vahab, Mikkel Thorup, and Morteza Zadimoghaddam. "Consistent Hashing with Bounded Loads." arXiv preprint arXiv:1608.01350 (2016).

[45] D. G. Thaler and C. V. Ravishankar. "Using name-based mappings to increase hit rates." IEEE/ACM Transactions on Networking (TON) 6.1 (1998): 1-14.

[46] Shvachko, Konstantin, et al. "The hadoop distributed file system." Mass storage systems and technologies (MSST), 2010 IEEE 26th symposium on. IEEE, 2010.

## APPENDIX A PSEUDO CODE OF ASURA

```
Input:
  datum_ID is ID of datum
  array_size is array size of segment_lengths and segment_numbers
  segment_lengths[] is array of segment_length, which is each segment.
  node_numbers[] is array of node number corresponding to each segment number.

Output:
  a segment number that is assigned to a data-storing node.

unsigned int asura(unsigned int datum_ID, unsigned int array_size, float segment_lengths[], int node_numbers[])
{
    c_max = DEFAULT_MAXIMUM_RANDOM_NUMBER;
    loop_max = 0;

    while(c_max < array_size){
        c_max *= 2;
        loop_max++;
    }

    initialize_control_variable_for_random_number_generator (control_variable_for_initialization, Datum ID);
    for(I = 0; I <= loop_max; i++){
        control_variable_is_used[i] = 0;
        seed_for_control_variables[i] = generate_integer_random_number (control_variable_for_initialization);
    }

    do{
        c = c_max;
        loop = loop_max;
        while(1){
            if(control_variable_is_used[loop] == 0){
                initialize_control_variable_for_random_number_generator (control_variables[loop], seed_for_control_variables[loop]);
                control_variable_is_used[loop] = 1;
            }
            do{
                result = generate_float_random_number_not_less_than_0_and_less_than_1 (control_variables[loop] ) * (float)c;
            }while(result >= maximum_segment_number_plus_1);

            c = c / 2;
            if(result >= c || loop == 0){
                break;
            }
            loop--;
        }
        int_result = (unsigned int)result;
    }while(int_result + segment_lengths[int_result] < result);

    return node_numbers[int_result];  // a node number that is assigned to a data-storing node
}
```

## APPENDIX B ORDER OF CALCULATION TIME OF ASURA

The expectation value of the number of random number generations does not depend on the number of nodes in ASURA. The proof is below.

This proof has the following presupposition.
- The number line starts from 0.0.
- Segment lengths are 1.0.
- There are no gaps between the segments.
- The following parameters are used:
  S: Initial maximum of random number
  α: Increase ratio of maximum random number
  n: Maximum segment number + segment length of the segment that has the maximum segment number
  h: Length of the hole that does not have corresponding node

First, the number of base random number generator, which is represented as x, to generate one ASURA random number is:

$$S\alpha^{(x-1)} < n \leq S\alpha^x$$
$$\log_\alpha (S\alpha^{(x-1)}) < \log_\alpha n \leq \log_\alpha (S\alpha^x)$$
$$x - 1 + \log_\alpha S < \log_\alpha n \leq x + \log_\alpha S$$
$$x - 1 < \log_\alpha \left(\frac{n}{S}\right) \leq x$$

Because x is an integral number,

$$x = \left\lceil \log_\alpha \left(\frac{n}{S}\right) \right\rceil \ldots (1)$$

The probability of generating an ASURA random number that points to a segment is

$$\frac{n-h}{S\alpha^x}\ldots(2)$$

By definition,

$$1 \leq \frac{S\alpha^x}{n} < \alpha$$

$$n \leq S\alpha^x < \alpha n$$

Thus,

$$\frac{n}{n-h} \leq \frac{S\alpha^x}{n-h} < \frac{\alpha n}{n-h}$$

$$\frac{1}{1-\frac{h}{n}} \leq \frac{S\alpha^x}{n-h} < \frac{\alpha}{1-\frac{h}{n}}\ldots(3)$$

The expected number of base random number generation to generate ASURA random number is:

$$\frac{S\alpha^x - S\alpha^{x-1}}{S\alpha^x} + 2\frac{S\alpha^{x-1} - S\alpha^{x-2}}{S\alpha^x} + \ldots + x\frac{S\alpha - S}{S\alpha^x} + (x+1)\frac{S}{S\alpha^x}$$

$$= 1 + \frac{1}{\alpha} + \frac{1}{\alpha^2} + \ldots + \frac{1}{\alpha^x}$$

$$= \frac{\left(\frac{1}{\alpha}\right)^{(x+1)} - 1}{\frac{1}{\alpha} - 1}$$

$$= \frac{\alpha^{x+1} - 1}{\alpha^{x+1} - \alpha^x}$$

$$= \frac{\alpha}{\alpha - 1} - \frac{1}{\alpha^x(\alpha - 1)}\ldots(4)$$

Then from Eqs. (2) and (4), the expected number of base random number generation to generate ASURA random number that points to a segment is:

$$\left(\frac{n-h}{S\alpha^x}\right)^{-1}\left(\frac{\alpha}{\alpha-1} - \frac{1}{\alpha^x(\alpha-1)}\right)$$

$$= \frac{S\alpha^x}{n-h}\left(\frac{\alpha}{\alpha-1} - \frac{1}{\alpha^x(\alpha-1)}\right)\ldots(5)$$

Here, we want to know the order of calculation when n increases. In Eqs. (5), $\frac{S\alpha^x}{n-h}$ only depends on $\frac{h}{n}$ from Eqs. (3), $\frac{\alpha}{\alpha-1}$ is constant, and $\frac{1}{\alpha^x(\alpha-1)}$ becomes 0 from Eqs. (1), when n increases. Thus, the maximum limitation of the expected number of base random number generation to generate ASURA random number that points to a segment only depends on $\frac{h}{n}$.

In conclusion, the expected number of base random number generation to generate ASURA random number that points to a segment approaches a constant when n is increasing, if $\frac{h}{n}$ is constant.